\documentstyle{article}

\def\Journal#1#2#3#4{{#1} {\bf #2}, #3 (#4)}

\def\PRD{{\em Phys. Rev.} D}

\def\al{\alpha}
\def\be{\begin{equation}}
\def\ee{\end{equation}}
 \def \dodo  { \partial ^0   \partial _0 }

\def \sig {\sigma}

\def \gam {\gamma}
\def \noi {\noindent}

\def \hron {{\cal H}}

\def \half  {  {1 \over 2 } } 
\def \eron {{\cal E}}
\def \sron {{\cal S}}

\begin{document}

\title{ISOMETRIC INVARIANCE OF THE POSITIVE-FREQUENCY KERNEL IN GENERALIZED 
FRW SPACETIMES}

\author{ Ph. DROZ-VINCENT   \\
      \medskip
\small Gravitation et Cosmologie Relativistes     \\
\small Universit\'e Pierre et Marie Curie  \\
\small  (CNRS, ESA 7065)                  \\
\small  T. 22-12 (boite courrier 142),      \\
\small  4 Place Jussieu, 75252 Paris Cedex 05, France}
\maketitle
\date{\ }
\begin{abstract}
We consider the Klein-Gordon equation with minimal coupling in FRW-like 
spacetimes, with compact space sections (not necessarily isotropic neither 
homogeneous). In such spaces, the bi-scalar kernel allowing to select the 
positive-frequency part of any solution is developed on modes, using the 
eigenfunctions of the three-dimensional Laplacian.
Of course this kernel is not unique, but any choice of it turns out to be 
manifestly invariant  under the isometries of space sections.  In the generic 
case (excluding  a special law of evolution for the scale factor),  spacetime 
has no nore symmetries (connected with the identity) than those inherited from 
space sections. Therefore, in this case, any choice of the kernel is also 
invariant under the {\em spacetime\/} isometries connected with identity. It 
is associated with a definition of hermitian scalar product that enjoys the 
same invariance properties, entailing that spacetime symmetries connected with 
the identity are unitarily represented.
Since all possible kernels are invariant and related one to another by a 
unitary transformation, we contemplate the possibility that all these 
definitions of the vacuum correspond to different representations with the 
same physical content.
\end{abstract}

\section{Quantization in curved spacetime}
\subsection{Isometric Invariance Principle}\label{subsec:isom}

\noi  Our goal is a theory of quantum {\em free} fields, in a given 
spacetime;   
before undertaking  the construction of Fock space we need to define the 
Hilbert space suitable for describing the motion of a single particle
\footnote{Dedicated to Lluis Bel on the occasion of the Spanish Relativity 
Meeting 1998}.
Therefore we consider here the Klein-Gordon (KG) equation 
\be   (\nabla ^2 + m^2 ) \Psi = 0               \label{eq:weq}    \ee
for a complex valued wave function $\Psi (x)$, describing the
 {\em minimal coupling}  of a scalar particle with gravity.
    The sesquilinear form 
\be  (\Phi ; \Psi) =  \int j^\mu (\Phi, \Psi)  \   d\Sigma _ \mu    
\label{eq:sesq} \ee
constructed from the Gordon current $ j^\nu (\Phi , \Psi) $
is conservative with respect to changes of hypersurface $\Sigma$ provided 
$\Phi $ and $\Psi$ are solutions to the KG equation. But it {\em is not} 
positive definite.  In order to exhibit a candidate for one-particle Hilbert 
space,   the linear space of solutions must be split in two subspaces.
  In one of them (further identified as positive-frequency space) the 
restriction of $(\Phi ; \Psi)$  must be definite  positive.

\noi The trouble is that, in nonstationary spacetimes, such splitting is
 not unique.
 Nevertheless, criteria for its  determination have been given soon,
either in terms of finding a real linear
 operator $J$ with $J^2 = - Id$, determining a complex structure in the space
 of real solutions ~\cite{seg},   
or alternatively  in terms of a  projector $\Pi ^+= \half (1 + i J)$
which projects any complex solution into the positive-frequency subspace.
Application of  $J$ or $\Pi ^+$ to a solution $\Psi$ is  carried out with the 
help of a kernel~\cite{ouch}, existence of which was proved
 by C. Moreno~\cite{mor}  under very general global assumptions.

\noi
  But in spite of the enormous amount of litterature up to now devoted to 
quantization in curved background,
 it seems that the role of spacetime isometries has not received 
all the attention it deserves ~\cite{bir},
 with an exception for de Sitter spacetime ~\cite{sit}.

\noi   
In view of the fundamental role played by Poincar\'e  group   with respect to
 quantum mechanics in Minkowski spacetime, 
we propose that quantization in any curved spacetime should satisfy this
 Principle:

\noi
 { \sl  Quantum mechanics  of free particles must be invariant under
 all spacetime symmetries  continuously connected with the Identity}.

\noi 
For simplicity we consider here only  continuous isometries, and postpone a 
discussion about discrete ones.

\subsection{The positive-frequency kernel}

The theory of retarded and advanced Green functions has been toroughly
 investigated for many years. These objects are unambiguously defined under 
very general assumptions (global hyperbolicity).

\noi          More problematic is  the kernel  $D^\pm$
 which allows  for   defining positive-frequency (resp. negative-frequency)
 solutions of the KG equation through the formula
\be  (\Psi ^\pm) (y) =  (D^\pm _y  ; \Psi ) = \int j^\al (D^\pm _y , \Psi) \  
d\Sigma  _\al      
\label{eq:defD}    \ee
where  $\Psi ^\pm  =  \Pi ^\pm    \  \Psi $
 is the positive-frequency part of $\Psi$,
 and $y$ is an arbitrary point of $V_4$.
Of course,  $D^\pm$  must satisfy the KG equation in both arguments;
the notation with subscript $y$ indicates that integration is 
performed on the variable $x$.
Kernel $D^+$ has  fundamental importance, for   
 the quantum field of the particle  must be defined as  anihilation 
operator   associated  with  the one-particle state  $D^+$.
By isometric  invariance we mean  
  $$ D^+  (x, y)   =  D ^ + ( Tx , Ty)     \label{eq:iso}         $$
for any metric-preserving transformation $T$ of the connected isometry group. 

\noi  When spacetime has the topology of ${\bf R} \times V_3$, with $V_3$ 
compact and connected,  Moreno's theorem ~\cite{mor} ensures 
that many $D^+$  actually exist (whatever may be the local form of the 
metric).  
Two such kernels are related by      a unitary transformation. 
 But  if   spacetime admits  Killing vectors, we cannot yet be 
satisfied with  any choice of $D^+$, before we answer  the question whether 
this kernel is  {\em invariant by action of the
isometries of} $(V_4, g)$. 
         This issue remains an open problem for arbitrary 
metric, but in the sequel  we focus on a class of spacetimes where isometries 
are under control and the KG  equation is separable.

\section{Quantization in Generalized FRW spacetimes}

\noi
We assume  that the spacetime manifold is 
$ V_4  = {\bf R } \times  V_3 $ with connected and compact $V_3$, 
and for some time scale $t$
\be      ds^ 2 = 
    B^ 6 dt^2 - B^ 2 d \sig ^ 2 
\label{eq:frw}   \ee
where  $B$ is a strictly positive function of $x^0 =t $,  and 
$ d \sig ^2 =  \gam _{ij}( x^ k ) dx^i dx^j  $
 defines an elliptic metric.
Notice that
  $ (V_3, \gamma ) $
may have  arbitrary curvature.     
The form (\ref{eq:frw}) provides a  generalization of 
 Friedmann-Robertson-Walker
 line element. Lorentzian manifolds  with such a metric have been 
systematically studied~\cite{carot}~\cite{sanch}. 
Beyond  the conventional FRW case, they can represent a  universe filled with a 
non-perfect fluid characterized by {\em anisotropic}  pressures~\cite{port}.   
They  have two nice properties:

\noi
--- In the generic case $(V_4, g)$  has no Killing vector beside trivial ones 
corresponding to the isometries of $(V_3, \gamma)$.
Cases where  additional Killing vectors arise require  exceptional laws  
of evolution for the scale factor~\cite{carot}~\cite{sanch}; they 
naturally   include the  de Sitter manifold.

\noi
---The KG equation is separable for special solutions associated with   
"modes"; the time dependence of mode solutions  is determined  by an ordinary
 diferential equation of second order. This situation stems from existence of
a certain  first integral, proportional to kinetic energy. 
Indeed  the three-dimensional Laplacian  $\Delta_3$ 
associated with metric $\gamma$ commutes with $\nabla ^2$, thus also  with 
the operator in l.h.s. of (\ref{eq:weq}).      This fact  permits 
 to reduce  the wave  equation into a one dimensional problem for 
time dependence, 
supplemented with an elliptic spectral problem  for space dependence. 
 
\subsection{Kernel at a given mode}
\noi
By extension of  the usual terminology we define 
  {\sl Mode} $n$  as the linear space  $\hron _n$
  of solutions to (\ref{eq:weq})  that are also 
eigenfunctions of  $-\Delta _3$  with eigenvalue  $\lambda _n$.   We  also 
refer to $\hron _n$   as "kinetic-energy shell". Separation of frequencies is 
supposed to respect the mode decomposition, therefore in each shell $\hron_n$ 
we look for  projectors   $\Pi^{\pm} _n$ associated with  kernels 
$D^{\pm} _n$.  Notice that $\hron_n$ is orthogonal to $\hron_l$ for $l \not= 
n$ in the sense of the sesquilinear form~\cite{droz}.

\noi   In this Section kinetic energy is kept fixed and the label $n$ 
referring to a determined eigenvalue $\lambda_n$ is provisionally dropped.

\noi       Let us  consider a mode solution, say $\Phi$. 
For some nonnegative $\lambda \in Spec(V_3) $ we have
$   \Delta_3  \Phi =  - \lambda \Phi   $
so  (1) reduces to
\be (\dodo +  \lambda  B^ {-2}  + \mu   )  \Phi  = 0   \label{eq:Reduc}  \ee 
with   $  \partial ^0  = B^{-6} \partial _0       $ and $\mu = m^2$.
The space variables $ x ^ j$ can be ignored in solving (\ref{eq:Reduc}) for 
$\Phi$. 
This equation   is always of second order.
        It is well-known~\cite{berg}   that  the eigenspace  
${\cal E}$  of  $- \Delta _3$  in $  C^\infty (V_3) $,  associated 
with the eigenvalue  $\lambda $  has finite dimension, say      $r$.

\medskip
\noi                      Let $ {\cal S} $ 
be the two-dimensional space  
of $C^ \infty $ complex-valued functions of  (a single variable)  $t$
 satisfying the equation
$  ( \dodo +  \lambda   B^ {-2}  + \mu   )  f = 0      $
and let  the functions 
$ f_{1 } (t) $ and 
$f _{2 } (t)  = f_1 ^*  $            form a basis of    $ {\cal S}   $. 
They respectively span  the one-dimensional subspaces 
$\sron ^{(1)} $ and $\sron ^{(2)} $.

\noi
We can always chose our notation and normalize  the basis of $\sron$  
according to the Wronskian condition 
$  W(f_1,  f_2) = -i $,  which amounts to  associate   $f_1, f_2$  respectively 
with  positive and negative frequencies. Call {\em admissible} such a basis.

\noi   With this convention we  proved, Propo.3 in [\cite{droz}],
 that the restriction of $ (\Phi  ;  \Phi)$   to  
 ${\cal S}^{(1)}   \otimes {\cal E} $ is positive definite (resp. negative,  
in  $ {\cal S}^{(2)}   \otimes {\cal E} $). 

\noi
The Hilbertian {\sl scalar product\/} is defined as
  $ <\Phi , \Psi >= \pm (\Phi ; \Psi) $ 
respectively in  $ \sron  ^{(1)} \otimes  \eron $  and 
        $ \sron  ^{(2)}  \otimes \eron $.
It crucially depends on  the splitting we have performed
(the choice of an admissible  basis) in $\sron$.      
  Notice that    $ \sron  ^{(1)}  \otimes \eron $               and 
        $ \sron  ^{(2)} \otimes  \eron  $                    are 
mutually orthogonal in the sense of this scalar product  as well as  in 
the sense of the sesquilinear   form  (\ref{eq:sesq}).

\medskip
\noi  
If $\Psi$ is a positive-frequency solution in mode $n$ we must have, restoring 
now the mode label
$$ ( (D^+_n)_y ;  \Psi ) =   \Psi (y)           \label{eq:deplus}           $$
where the notation  $  (D^+_n)_y $ indicates that
 $D^+ _n$ is considered as a function of $x$ which additionally depend on $y$.
Let  $ E_{1,n},....E_{r,n} $ be a real  orthonormal  basis of $\eron_n$.
Use the notation
 $x= (t, \xi), \quad  y = (u, \eta) $ with $\xi , \eta  \in V_3$.   

\noi       The only possibility for $D^\pm  _n$  takes on the standard form 
\be                   D ^+_n  (y,x)  =
  f_{1,n} ^*  (u)  f _{1,n}    (t)    \Gamma_n (\eta, \xi)
                    \label{eq:kernl}  \ee
where the expression 
$$  \Gamma _n (\eta, \xi)  =   \sum E_{a,n}  (\eta) \    E _{a,n}   (\xi) 
   \label{eq:gamoto}  $$
is real and doesnot depend on the choice of  a real orthonormal basis
 in $\eron_n$.
It is intrinsically  determined by the spectral properties of $V_3$.

\noi It is straightforward  to check that  expression (\ref{eq:kernl})
 of $D^+_n$ actually  satisfies  equation (\ref{eq:defD}) as it should. 
    The only arbitrariness  in formula (\ref{eq:kernl}) is in the factor 
$f_{1,n} ^* (u)  f_{1,n} (t)  = f_{2,n} (u)  f_{1,n}  (t)$ 
 which depends on the choice of  an admissible  basis 
  in the two-dimensional space $\sron _n$.

\noi For usual FRW spacetimes
 ($V_3$ is of constant curvature), a basis of $\eron_n$ can be found in the 
literature~\cite{berg}.

\bigskip
 \noi     Let us now turn to isometric invariance.
If   $T$ is  an isometry {\em of the spatial metric}  $\gam _{ij}$, it acts on 
functions according to           $T F = F(T \xi)$.
Invariance of  $\Delta _3$  entails that  each   eigenspace  ${\cal E }_n $
 is globally invariant.
Moreover $T$ leaves invariant the  three-dimensional scalar product
$  ((F,G))  $.
Thus  $ E_{1,n} (T \xi) , .... E_{r,n} (T \xi ) $ 
is another real orthogonal basis  of $ {\cal E }_n$.  Finally 
$ \Gamma _n (T \eta , T \xi ) =   \Gamma _n  (\eta , \xi)    $ and we can 
write
$ D^+_n   (Ty , Tx) =  D^+ _n  (y,x)        $.

\noi
In the generic case  all isometries of $V_4$ are lifts of spatial isometries, 
so  we summarize:

\noi                          {\sl The only kernel $D^+ _n $ 
solution of the wave equation (1), eigenfunction of $-\Delta _3$ for eigenvalue 
$\lambda _n$
 and satisfying  (\ref{eq:defD}) where $\Psi$ is a  solution   in mode $n$, 
  is given by (\ref{eq:kernl}) and defined up to a change of admissible  
basis in the two-dimensional complex space $\sron _n$ (Bogoliubov 
transformation).
 In the generic case, it is isometrically invariant.}
                                                             
\bigskip
\subsection{ Sum over modes}

\noi  
Define  $ \hron ^+ =  \bigoplus  \sron _n ^{(1)}  \otimes  \eron _n $.
\noi                    The infinite sum
$    \displaystyle   \Phi = \sum  ^\infty     \Phi _n $
where  $\Phi _n  \in   \hron _n $,
always  exists in the distributional sense, 
if we define as {\sl test functions\/} 
the  sums   $\Psi = \sum \Psi _n $ having   an   arbitrary but 
{\it finite\/} number of terms  (terminating sums). This 
definition is invariant under spatial isometries.
In this sense we can assert:

\noi   {\sl The only kernel solution of the wave equation, mode-wise defined 
 and satisfying (\ref{eq:defD}) is given by  $ \displaystyle D^+ = \sum 
^\infty  D ^+ _n $.
It is defined up to a unitary  transformation
 $U = \bigoplus U_n$  where $U_n$ is an arbitrary Bogolubov 
transformation in mode $n$.
 In the generic case, $D^+$ is invariant  under the continuous  
isometries of $V_4$.}

\section {Conclusion}
\noi  As expected from the work of several authors,
the positive-frequency kernel is defined up to a unitary transformation. 
For generalized FRW spacetimes  we have additionally proved  that this  kernel 
is invariant under all spacetime isometries connected with the identity, 
{\em except} perhaps for very special forms of the scale-factor evolution. 
de Sitter manifold is one such exception. This does not contradict existence 
of a de Sitter invariant vacuum, but rather indicates that a mode-wize 
definition of $D^+$ is no longer satisfactory when the separation of space from 
time fails to be unique. 

\noi Returning  to the generic case, it can be easily read off from 
(\ref{eq:kernl})  that the connected group of spacetime isometries acts 
 unitarily in   $  \hron  ^+ $ whatever is the choice of admissible  basis 
made in each $\sron _n$.
 This fact strongly suggests to consider that all isometrically invariant 
  definitions of the {\em one-particle} sector  are equivalent representations
 of the same physics.

\noi
When expansion is statically bounded in past and future, this point of view 
amounts to   by-pass the  customary  "in and out" vacua  in favor of a unique 
 class of observers, not  submitted to asymptotic conditions.
This scheme  {\em at least} provides a reasonable  approximation  insofar
 as the curvature gradient in $V_3$ remains small and 
the scale factor does not vary too rapidly.

\noi  
Notice that in some particular cases,
 among all equivalent definitions, a distinguished vacuum  may
 arise after all,  
  in agreement with recent works ~\cite{droz}~\cite{bel}.


\end{document}